\documentclass[superscriptaddress,reprint,amssymb,amsmath,nobibnotes,aps]{article}
\usepackage{graphicx}
\usepackage{epstopdf}
\usepackage{authblk}
\usepackage{color}
\usepackage{subfig}
\usepackage{bm}
\usepackage{aas_macros}
\usepackage{appendix}
\usepackage{mathrsfs}
\usepackage{amsmath}
\usepackage[margin=.80in]{geometry}
\usepackage{multicol}

\newcommand{\bicep}{B{\sc icep}}
\newcommand{\PB}{POLARBEAR}

\newcommand\authemail[1]{%
  \begingroup
  \renewcommand\thefootnote{}\footnote{#1}%
  \addtocounter{footnote}{-1}%
  \endgroup
}

\begin{document}
     
\title{Precision Tests of Parity Violation Over Cosmological Distances}

\author{Jonathan P. Kaufman$^*$}
\author{Brian G. Keating}
\affil{Department of Physics, University of California, San Diego, 9500 Gilman Drive, La Jolla, CA 92093-0424, USA}
\author{Bradley R. Johnson}
\affil{Department of Physics, Columbia University, New York, NY 10027, USA}

\date{\today}

\maketitle


\begin{abstract}

Recent measurements of the Cosmic Microwave Background (CMB) $B$-mode polarization power spectrum by the \bicep2 and \PB\ experiments have demonstrated new precision tools for probing fundamental physics.  Regardless of origin, the detection of sub-$\mu$K CMB polarization represents a technological \emph{tour de force}. Yet more information may be latent in the CMB's polarization pattern. Because of its tensorial nature, CMB polarization may also reveal parity-violating physics via a detection of cosmic polarization rotation. Although current CMB polarimeters are sensitive enough to measure one degree-level polarization rotation with $>5\sigma$ statistical significance, they lack the ability to differentiate this effect from a systematic instrumental polarization rotation.  Here, we motivate the search for cosmic polarization rotation from current CMB data as well as independent radio galaxy and quasar polarization measurements. We argue that an improvement in calibration accuracy would allow the unambiguous measurement of parity- and Lorentz-violating effects.  We describe the CalSat space-based polarization calibrator that will provide stringent control of systematic polarization angle calibration uncertainties to $0.05^\circ$ -- an order of magnitude improvement over current CMB polarization calibrators.  CalSat-based calibration could be used with current CMB polarimeters searching for $B$-mode polarization, effectively turning them into probes of cosmic parity violation, ``for free" -- i.e. without the need to build dedicated instruments.

\end{abstract}

\section{Introduction} \label{sec:intro}

\authemail{$^*$E-mail: jkaufman@physics.ucsd.edu}

Lorentz invariance and invariance under CPT operations undergird all of modern physics: the laws of physics are not specific to our reference frame, and the evolution of a physical system is invariant under a simultaneous charge conjugation (C) and parity (P) and time (T) reversal. To the best of our abilities to test these assumptions, they remain intact.  Though the standard model respects \emph{CPT} symmetry, it manifestly violates \emph{Parity} conservation in the Weak sector, as demonstrated in the asymmetric angular distribution of electron emission in Cobalt-60 predicted by Lee and Yang in 1956 and measured by Wu et al. in 1957 \cite{LeeYang1956, Wu1957}. Furthermore, the compound CP symmetry can also be violated \cite{Christenson1964}. 
The unification of electromagnetism with the Weak force at high energies suggests that at sufficiently early cosmological epochs, electromagnetism may violate parity symmetry as well.  This hypothesis motivates the exploration of the consequences of parity violation on distant cosmic sources. 

Adding a Lorentz- and CPT-violating Chern-Simons interaction term to the electromagnetic Lagrangian rotates the polarization plane of linearly polarized\footnote{Circular polarization is not produced by the Thomson scattering mechanism responsible for the CMB's polarization, however \cite{Alexander2009, Finelli2009} have shown circular polarization can arise via coupling of the photons to an external vector field due to a Chern-Simons like interaction.} electromagnetic radiation propagating through vacuum \cite{Carroll1990}.  This rotation depends on optical path length such that the greater the distance the photons travel, the larger the amount of polarization rotation.  Extremely small deviations from the standard model's null-expectation can accrue to observable values over the millions, or even billions, of parsecs that the photons travel.  

Studying the polarization of distant galaxies is a useful probe of cosmic polarization rotation, however these measurements can be limited by model dependencies as described below.  Rotation of the Cosmic Microwave Background's (CMB) polarization offers a more sensitive probe as it is the most distant possible polarized source.  More importantly, CMB model uncertainties are relatively small and a robust detection is limited by the accuracy of the polarization angle calibration source\footnote{Up to the cosmic variance limit.  Faraday rotation can be a systematic contamination, however due to the strong frequency dependence of Faraday rotation, this can be decoupled via a multi-frequency measurement.}.

Unfortunately, the calibration of CMB detector polarization angles is significantly more complicated than calibrating flux responses \cite{Miller2009}.  Whereas planets, the CMB dipole, and even the temperature power spectrum can be used to calibrate absolute temperature response \cite{PlanckCal2013,Page2003,Reichardt2012}, there is no analogous standard for polarization angle calibration.  Self-calibration of detector polarization angles can be obtained using the CMB itself, however this comes at the expense of measuring cosmic polarization rotation \cite{KSY2013}. There are a paucity of well-characterized polarized astronomical sources that can be used as calibrators.  Furthermore the few astronomical sources that do exist suffer from frequency dependence, time variability, and are not visible from all observatories.  Hardware calibrators are difficult to fabricate and employ in the field.  Although polarization angle calibration uncertainty is degenerate with cosmic polarization rotation \cite{KSY2013,Yadav2013, Kaufman2013}, it is not at present a significant contaminant of $B$-mode polarization.  Current calibration uncertainties lead to errors far below the $B$-mode polarization at large or small scales as measured by \bicep2 and \PB\footnote{Though if $r$ were found to be significantly lower than the \bicep2 result, better polarization angle calibration would be required.} \cite{B22014,PB2014}.

\subsection{Historical Motivation}

In the early 1960s, engineers at Bell Labs used the 6 meter Holmdel microwave horn-reflector antenna as part of the Echo satellite communication system.  During commissioning, a team led by Ed Ohm thoroughly evaluated the systematic contributions of each component of the receiver.  After the systematic error budget was tabulated, there remained an unexplained 3.3 K deviation from the predicted antenna temperature.  In Ohm's words, this was attributed to the fact that ``the `$+$' temperature possibilities of [the receiver temperature uncertainties] must predominate" \cite{Ohm1961}.  Doroshkevich and Novikov cited Ohm's paper in a 1964 review of cosmic radiation predicted in Gamow's theory of the origin of the universe \cite{Doro1964}. Perhaps due to a misinterpretation of the 3.3 K excess, Doroshkevich and Novikov attributed the excess temperature to emission from the Earth's atmosphere. In actuality it was a residual signal persisting even \emph{after} atmospheric emission had been removed.

To definitively measure the source of the systematic excess emission, Penzias and Wilson built a Dicke switched radiometer for the same 6 meter Holmdel antenna which featured an accurate liquid Helium cooled reference load. This calibration load reduced systematic errors far below the levels Ohm obtained \cite{RadioAstro1983}.  With this improved calibration, they were able to convincingly determine that the excess was actually the CMB itself.

Cosmic polarization rotation is also important for testing general relativity through the Einstein equivalence principle (EEP).  In 1960, Schiff conjectured that any Lorentz-invariant theory of gravity that obeys the weak equivalence principle also obeys the EEP \cite{Schiff1960}.  However, in 1977 Ni found that the coupling of a pseudo-scalar field to electromagnetism would obey the weak equivalence principle but violate the EEP \cite{Ni1977}.  This coupling rotates the polarization plane of linearly polarized light \cite{Carroll1991} and thus constraints on cosmic polarization rotation also test the EEP and therefore gravity itself.

The field of cosmic polarization rotation measurements is also replete with claimed detections and misinterpretations. In 1997, a detection of cosmological polarization rotation was claimed from radio galaxy polarization measurements \cite{NR1997}, however this was quickly and definitively refuted \cite{Carroll1997}. Since 2010, several papers have claimed low significance evidence for polarization rotation by combining data from several CMB polarization experiments \cite{Xia2010, Xia2012, Li2014}.  However inaccuracies in polarization calibration prevent decoupling any cosmic polarization rotation from instrumental systematics.  Assessing the joint likelihood of these experiments is not trivial due to the hidden covariances in their calibration methods and systematic uncertainties in each experiment's polarization angle calibration.

\subsection{Outline}

A definitive measurement of polarization rotation of cosmic origin would be the first detection of the violation of CPT and Lorentz invariance. It would also test the Einstein equivalence principle and have dramatic consequences for cosmology and much of modern physics. Because of its potentially revolutionary impact, any measurement claim must be carefully scrutinized for the presence of systematic errors. We will demonstrate that, for the first time, we can obtain measurements of cosmic polarization rotation that are not limited by systematics, but rather by statistical errors alone. 

This paper is structured as follows: In Section 2, we discuss the theory behind cosmic polarization rotation and summarize constraints on polarization rotation from radio galaxies, quasars, and CMB polarization measurements.  We discuss these constraints in Section \ref{sec:discussion} and provide a path for improving these measurements in Section \ref{sec:forward} using a dedicated polarization angle calibration source called ``CalSat."  Our conclusions are in Section \ref{sec:conclusions}.


\section{Cosmic Polarization Rotation}\label{sec:CPR}
We begin by describing the physics of electromagnetic wave propagation in vacuum with and without cosmic polarization rotation. The standard model electromagnetic Lagrangian can be expressed as:
\begin{equation}
\mathcal{L}_{EM} = -\frac{1}{4} F_{\mu\nu}F^{\mu\nu},
\end{equation}
where $F_{\mu\nu}$ is the electromagnetic tensor, $F_{\mu\nu} = \partial_\mu A_\nu - \partial_\nu A_\mu$, and $A_\mu$ is the electromagnetic four-potential.  This Lagrangian leads to electromagnetic fields which respect CPT symmetry operations.  However, if we add the Chern-Simons Lagrangian, both CPT symmetry and Lorentz invariance will be broken.  The Chern-Simons Lagrangian is
\begin{equation}
\mathcal{L}_{CS} = - \frac{1}{2} p_\mu A_\nu \tilde{F}^{\mu\nu},
\label{eq:CS}
\end{equation}
where $p_\mu$ is the coupling four-vector.  This coupling four-vector creates a preferential direction in space-time, violating Lorentz invariance.  The spatial components of $p_\mu$ violate rotational invariance and the time component prevents invariance under a Lorentz boost. We can think of $p_\mu$ as a mass term, where $p_\mu p^\mu \equiv -m^2$.

The addition of this term to the electromagnetic Lagrangian is equivalent to modifying the four-current  $J^\nu \rightarrow J^\nu + p_\mu \tilde{F}^{\mu\nu}/4\pi$.  Setting $c$=1, following \cite{Carroll1990}, the solution of the wave equations modifies the dispersion relation from $\omega = k$ to
\begin{equation}
\omega^2-k^2 = \pm \left(p_0 k - \omega p \cos{\theta} \right) \left[1 - \frac{p^2\sin^2{\theta}}{\omega^2 - k^2}\right]^{-1/2},
\label{eq:disp_rel}
\end{equation}
where $p_\mu \equiv (p_0, \vec{p})$ is the coupling four-vector from the Chern-Simons Lagrangian, $p \equiv |\vec{p}|$, and $\theta$ is the angle between the spatial coupling vector $\vec{p}$ and the wave vector $\vec{k}$.  The $+ ~(-)$ corresponds to left-handed (right-handed) circularly polarized light.  
If $p_\mu$ is small, we can Taylor expand Equation \ref{eq:disp_rel}.  To first order, we obtain:
\begin{equation}
k = \omega \mp \frac{1}{2} \left( p_0 - p\cos\theta \right).
\label{eq:kCB}
\end{equation}

\subsection{Optical Polarization Rotation}

A linearly polarized photon with polarization angle $\phi$ can be represented as $| \psi(\phi) \rangle $.  This photon can be decomposed into left-handed and right-handed circularly polarized states:
\begin{equation}
| \psi(\phi)\rangle = | \psi_R\rangle + e^{i2\phi} | \psi_L\rangle.
\end{equation}
If this photon propagates through a medium exhibiting optical activity (also called. ``circular birefringence"), one circular polarization state will lag the other and the linear polarization axis will rotate by an angle
\begin{equation}
\alpha = \frac{1}{2} (k_R - k_L) L,
\label{eq:alpha_n}
\end{equation}
where $k_R$ ($k_L$) is the wavenumber for right-handed (left-handed) circularly polarized light and $L$ is the optical path length of propagation through the medium.  This effectively creates a different index of refraction for each polarization state.
From Equations \ref{eq:kCB} and \ref{eq:alpha_n}, we can write the polarization rotation as
\begin{equation}
\alpha = -\frac{1}{2} \left(p_0 - p  \cos{\theta}\right) L.
\label{eq:polrot}
\end{equation}
This rotation angle $\alpha$ affects the polarization of radio galaxies and the CMB in unique ways.

\subsection{Polarization of Extragalactic Sources} 

Radio galaxies can be polarized due to synchrotron emission, and if the orientation of the magnetic fields with respect to the spatial morphology of the galaxy is known, the polarization alignment can be inferred.  Many of these radio sources are elongated along a particular axis\footnote{Since the important quantity is the difference between this angle $\psi$ and the polarization axis, the reference to which $\psi$ is measured is not important, so long as it is consistent with the polarization axis reference.} in space, $\psi$.  Since the emission mechanism is synchrotron radiation, the polarization axis is expected to either align with this elongation axis or be perpendicular to it.  Thus the measured angle $\chi$ should be either
\begin{eqnarray}
\chi & \approx & \psi \nonumber\\
 &\text{or} & \nonumber\\
\chi & \approx & \psi + 90^\circ.
\label{eq:RG}
\end{eqnarray}
Polarization rotation would be observable by measuring a difference between $\psi$ and $\chi$:
\begin{eqnarray}
\alpha & = & \chi - \psi \nonumber\\
&\text{or} & \nonumber\\
 \alpha - 90^\circ & = & \chi - \psi.
 \label{eq:alphaRG}
\end{eqnarray}
This method however must be corrected for Faraday rotation and projection effects \cite{Carroll1997}, and it is dependent on inhomogeneities in the host galaxy's magnetic fields.  The relationship in Equation \ref{eq:RG} is true only for a statistical ensemble of identical sources; any given source may not obey the relationships in Equations \ref{eq:RG} and \ref{eq:alphaRG} in the absence of cosmic polarization rotation.

In addition to their radio wavelength polarization properties, the polarization of the ultraviolet emission of high redshift galaxies can be used to constrain polarization rotation.  In this method, the UV polarization axis of a distant galaxy is expected to be perpendicular to its spatially elongated axis due to scattering of the central quasar's radiation, as stated by the unified active galactic nuclei model \cite{Antonucci1993}. Any deviations from this expected polarization orientation can be measured, similar to the radio galaxy method.  

Limits on polarization rotation derived from these methods have constrained cosmic polarization rotation as shown in Table \ref{table:RG_CB} and Figure \ref{fig:alphas} \cite{Carroll1998,UV11997,UV22010,Kamionkowski2010}.

\begin{table*}[h]
\caption{Rotation Angles derived from radio-wave and ultraviolet observations of extragalactic source polarization, as shown in \cite{RGCB2010}.  Note that we use the CMB sign convention as discussed in \cite{Galaverni2014}.}
\centering
\begin{tabular}{|c|c|c|c|}
\hline Method & Redshift & $\alpha$ (degrees)\\
\hline Radio &$ z < 2$ & $0.6 \pm 1.5$ \cite{Carroll1998,Leahy1997}\\
\hline UV & $z = 0.811$ & $1.4 \pm 1.1$ \cite{UV11997}\\
\hline UV & $ 2.2 < z  < 3.6$ & $ 0.8 \pm 2.2$ \cite{UV22010}\\
\hline
\end{tabular}

\label{table:RG_CB}
\end{table*}

If cosmological polarization rotation is a small effect, these methods may not have sufficient constraining power above the intrinsic model uncertainties described above.  From Equation \ref{eq:polrot}, we see that polarization rotation is dependent on optical path length.  We can convert this to redshift dependence:
\begin{equation}
\alpha = -\frac{p_0 - p \cos{\theta}}{3H_0}\left[ 1 - (1+z)^{-3/2} \right],
\label{eq:redshift}
\end{equation}
where $H_0$ is the Hubble constant and $z$ is redshift \cite{Carroll1990}.  Higher redshift sources increase the sensitivity to polarization rotation, motivating the search for more distant polarized sources.  CMB polarization is the most distant polarized object, produced at redshift of $z \sim 1100$, and is therefore the most sensitive probe of cosmic polarization rotation. However, from Eq. 10, we note that the difference in the CPR accumulated from $z=3.6$ to that accumulated from z=1100 is only $\sim 10\%$.  If CPR exists, it would be detectable with both methods, providing a valuable cross-check from methodologies with completely different systematic susceptibilities. The statistical precision of modern CMB polarimeters is more than adequate to constrain or detect minute levels of cosmic polarization rotation. However, such instruments lack sufficiently accurate calibration sources, as we will outline below.


\subsection{Constraints from the CMB}

Having demonstrated the limitations of radio and UV galaxy measurements for constraining cosmic polarization rotation, we turn to the CMB: the most distant source of electromagnetic radiation.
The statistical properties of the CMB are characterized in terms of its angular power spectra, $C^{XX'}_\ell$, where
\begin{equation}
C_{\ell}^{X X'} \delta_{\ell \ell'} \delta_{m m'} = \left\langle a_{\ell m}^{X*} a_{\ell' m'}^{X'} \right\rangle.
\end{equation}
Here, $X$ and $X'$ can be any combination of temperature anisotropy (``$T$"), $E$-mode, or $B$-mode polarization, and $a_{\ell m}$ is the amplitude of a given $\ell$ and $m$ mode of the temperature or polarization maps decomposed into their spherical harmonics \cite{Seljak97,KKS1997}.

The standard $\Lambda$CDM cosmological model predicts autocorrelations of the temperature anisotropy, $E$-mode polarization, and $B$-mode polarization, if gravitational lensing or Inflationary gravitational waves are included.  Correlations between temperature anisotropy and $E$-mode polarization are also expected.  The Standard Model does not predict correlations of temperature with $B$-mode polarization, or between $E$-mode and $B$-mode polarization.  That is, it predicts $C^{TB}_\ell =C^{EB}_\ell = 0$.

However, if there is a cosmic polarization rotation as in Equation \ref{eq:polrot}, there will be $E$-mode and $B$-mode mixing that will generate $TB$ and $EB$ correlations \cite{LueWangKamion1999}:
\begin{eqnarray}
C_{\ell}^{'TB} &=& C_{\ell}^{TE} \sin(2\alpha) \nonumber \\
C_{\ell}^{'EB} &=& \frac{1}{2}\left( C_{\ell}^{EE} - C_{\ell}^{BB} \right) \sin(4\alpha),
\label{eq:TBEB}
\end{eqnarray}
where $\alpha$ is the polarization rotation angle \cite{Feng2006}.
Thus measuring $TB$ and $EB$ deviations from null is a test of polarization rotation, whether cosmological or instrumental in origin\footnote{While $EE$ and $BB$ can the measure magnitude of polarization rotation, only $TB$ and $EB$ can constrain the \emph{sign} of the rotation.} \cite{KSY2013}.

So far, we have focused on isotropic polarization rotation; however inhomogeneities in the coupling scalar field can lead to direction-dependent, or anisotropic, polarization rotation \cite{Li2008,Gluscevic2009,Alighieri2014}.  In addition, $TB$ and $EB$ correlations can be generated by chiral gravity \cite{CMS2008}, as well as primordial magnetic fields \cite{PMF2012} and cosmic defects \cite{Moss2014}.

Several CMB experiments have constrained polarization rotation using $TB$ and $EB$ correlations.  These are shown in Table \ref{table:CMB_CB}, and several authors have attempted to co-add results from different experiments \cite{Xia2010, Xia2012, Li2014, Galaverni2014}.  However, no detection, or even meaningful joint-constraints, can be claimed from the naive co-addition of the central values and uncertainties, as all data sets are dominated by the large systematic uncertainties from polarization angle mis-calibration effects which do not average down across experiments. In fact, with systematic errors much greater than statistical errors, it is fundamentally impossible to simply add the individual constraints together.  What is needed is a series of measurements, each of which has systematic errors far subdominant to their statistical errors.  To detect cosmic polarization rotation we require more precise polarization angle calibration.

\begin{table*}[h]
\caption{Rotation angles derived from the polarization of the CMB along with their statistical and systematic uncertainties (in parentheses where available).}
\hspace*{-0.2cm}
\begin{tabular}{|c|c|c|c|c|}
\hline Experiment & Frequency (GHz) & $\ell$ range & $\alpha$ (degrees) & calibration method\\
\hline WMAP7 \cite{Komatsu2011}& 41+61+94 & 2 - 800 & $-1.1 \pm 1.4 ~(\pm 1.5)$ & pre-launch polarized source/Tau A\\
\hline BOOM03  \cite{BOOM03}& 143 & 150 - 1000  & $-4.3 \pm 4.1$ & pre-flight polarized thermal source\\
\hline QUaD \cite{Wu2009}& 100 & 200 - 2000 & $-1.89 \pm 2.24 ~(\pm 0.5)$ & polarized thermal source\\
\hline QUaD \cite{Wu2009} & 150 & 200 - 2000 & $+0.83 \pm 0.94 ~(\pm 0.5)$ & polarized thermal source\\
\hline \bicep1 (DSC) \cite{Kaufman2013} & 100+150 & 30 - 300 & $-2.77 \pm 0.86 ~(\pm1.3)$ & dielectric sheet\\
\hline \bicep1 (grid) \cite{Kaufman2013}& 100+150 & 30 - 300  & $-1.71 \pm 0.86 ~(\pm 1.3)$ & polarized microwave source\\
\hline \bicep1 (design)  \cite{Kaufman2013}& 100+150 & 30 - 300  & $-1.27 \pm 0.86 ~(\pm 1.3)$ &``as-designed"\\
\hline \PB \cite{PB2014}& 150 & 500 - 2100 & $ -1.08 \pm 0.20~(\pm 0.5)$ & Tau A\\
\hline \bicep2  \cite{B22014,B2Instr2014,Aikin_thesis}& 150 & 30 - 300  & $\alpha  \approx -1 \pm 0.2 ^a$ &dielectric sheet\\
\hline ACTPol \cite{ACTPol2014} & 146 & 500 - 2000 & $\alpha = -0.2 \pm 0.5^b$ & ``as-designed"\\
\hline
\end{tabular}

\vspace{.15 in}
$^a$\footnotesize Although \bicep2 has a precise statistical constraint on the rotation angle, the reported angle's systematic accuracy has not been published.  A more thorough investigation of the systematic uncertainty, in the style of \cite{Kaufman2013} is necessary.\\
$^b$\footnotesize ACTPol reports a $1.0^\circ \pm 0.5^\circ$ offset between the CMB-derived rotation angle and the Tau A derived angle, consistent with \PB's measurements.  The rotation angle may actually be as large as $-1.2^\circ$, depending on whether Tau A is used as the absolute polarization angle reference.

\label{table:CMB_CB}
\end{table*}


\section{Discussion} \label{sec:discussion}

\begin{figure*}[h]
\centering
\includegraphics[width=0.55\textwidth]{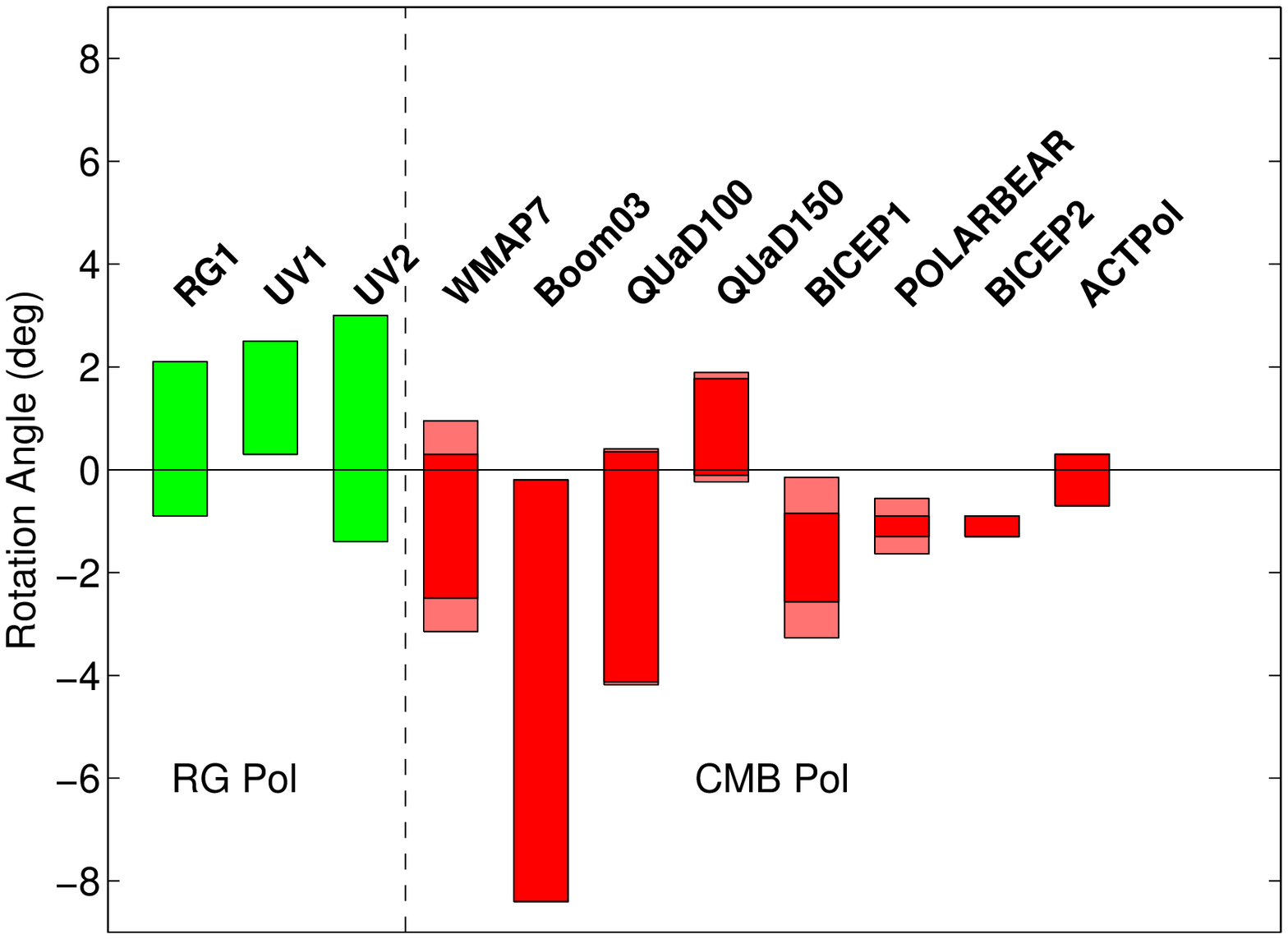}

\caption{Radio, UV, and CMB constraints on cosmic polarization rotation.  Radio galaxy polarization derived angles are plotted in green and CMB polarization derived angles are plotted in red.  The  boxes represent the range of rotation angles with their statistical uncertainty.  The black boxes show the total uncertainty, defined as the statistical uncertainty added in quadrature with systematic uncertainty (where available).} 
\label{fig:alphas}
\end{figure*}

We show constraints from three independent astrophysical methods: galaxy polarization in the radio spectrum and UV spectrum, and CMB polarization in Figure \ref{fig:alphas}.  These have constrained polarization rotation to be a small effect and are all limited by their relatively large systematic error contributions.

The only CMB instrumental error capable of producing a signature consistent with an overall polarization rotation is a miscalibration of detector polarization orientations\footnote{We note that any birefringence in the optical elements cannot mimic cosmic polarization rotation since the calibration source and the CMB radiation pass through the same optical chain.} \cite{Miller2009}.  For example, no other instrumental error will produce a polarization rotation angle that is the same for both the $TB$ and $EB$ estimators.  To date, several calibration methods have been employed, which we briefly review here. 

\begin{itemize}

\item The \bicep1 and \bicep2 experiments use a dielectric sheet calibrator located in the near field to produce a small polarized signal from unpolarized atmospheric and ambient radiation \cite{Kaufman2013,Takahashi2010,B2Instr2014}.  \bicep2's absolute polarization calibration used the self-calibration technique using the CMB's $TB$ and $EB$ spectra \cite{KSY2013}. \bicep2's polarization calibration was more than adequate to constrain $r=0.2$. \bicep1 also employed a broad-band noise source combined with a wire grid located in the far-field.  To observe this source, \bicep1 required a mirror oriented at 45$^\circ$ above the instrument, reflecting the beams to elevations closer to the horizon. 

\item SPTpol employed a similar broad-band source to obtain polarization calibration to approximately $1^\circ$ but it was located in the near-field, short of the Fraunhofer distance \cite{George2012}.  

\item WMAP measured polarization angles prior to launch using near-field polarized microwave sources \cite{Page2003}.  

\item QUaD employed a near-field polarizing wire grid, similar in principle to the dielectric sheet method \cite{Hinderks2009}. 

\item \PB\ used observations of Tau A at 90 GHz extrapolated to 150 GHz, and also employed self-calibration using the CMB's $EB$ spectrum \cite{PB2014}. 

\item ACTPol used modeling coupled with pointing observations to obtain their ``as-designed" angles, which they then compared to CMB-derived and Tau A-derived angles \cite{ACTPol2014}.  \bicep1 also employed as-designed angles but with a less complicated optical model (due to the lack of a reflector).  

\item The Planck polarization power spectra are expected to be released by late-2014.  The Planck polarization angle systematic uncertainty is expected to be $\simeq0.3^\circ$ based on pre-flight near-field wire grid calibration measurements \cite{Rosset2010,Pajot2010}.

\end{itemize}

A list of current calibrators and their uncertainties is in Table \ref{table:cal}. These methods have provided adequate polarization angle calibration to constrain gravitational waves if $r\geq0.01$, to measure gravitational lensing $B$-modes \cite{PBlens}, to measure $B$-mode polarization from the CMB alone \cite{PB2014} as well as to measure $B$-mode polarization in cross-correlation with data from Herschel \cite{Hanson2013, PBxcorr}. However, these calibration methods are inadequate for disentangling the effects of sub-degree cosmic polarization rotation from instrumental polarization rotation. These hardware and astrophysical sources suffer from a panoply of systematic effects, briefly summarized here. 

Many effects are attributable to hardware calibrators being placed in the antenna's near-field. Diffraction effects may preclude the accurate extrapolation of polarization angles from the antenna near-field (where calibration is performed) to the far-field (where the celestial signals originate). The dielectric sheet calibrator method is in the near-field and is susceptible to installation error and variations in the sheet thickness.  The far-field wire polarized microwave source suffers from alignment errors, polarizing wire-grid imperfections, and non-idealities in the reflector used to couple the beam to the telescope\footnote{It is entirely possible these combined effects are well below the quoted systematic error, however a more thorough analysis of the calibrator is needed.}.  \PB\ and ACTPol's systematic uncertainty is limited by the $0.5^\circ$ systematic uncertainty on Tau A \cite{Aumont2009}. Tau A is not visible from the experiments at the South Pole, though Cen A has been measured by QUaD in an attempt to obtain an astrophysical source visible from Polar observatories \cite{Zemcov2009}. A cross-check of the Tau A-derived polarization angles with those obtained from Cen A was performed in \cite{PB2014}. The two celestial sources were found to be in agreement, though the precision with which Cen A has been measured precluded an improvement upon the Tau A-derived value.  \bicep1, \bicep2, and \PB\ apply ``self-calibration" from the CMB's $TB$ and $EB$ spectra to de-rotate the offset from their polarization angles \cite{Kaufman2013,B22014,PB2014}, which precludes the detection of cosmic polarization rotation.  

\begin{table}[h]
\caption{Current calibration methods and their precision.}
\centering
\begin{tabular}{|c|c||c|c|}
\hline Hardware Method & Precision & Celestial Source & Precision \\
\hline dielectric sheet & $1.3^\circ$ \cite{Kaufman2013} & Tau A & $0.5^\circ$ \cite{PB2014}\\
near-field wire grid & $1.7^\circ$ \cite{Hinderks2009}& Cen A & $1.7^{\circ a}$ \cite{Zemcov2009}\\
as-designed & $0.5^\circ$ \cite{ACTPol2014}& &\\
rotating far-field source & $<1^\circ$ \cite{George2012}& &\\
\hline
\end{tabular}

\vspace{.15 in}
$^a$\footnotesize Note that the precision with which Cen A has been measured is limited by systematic errors in the QUaD wire grid calibration scheme.

\label{table:cal}
\end{table}


\section{Path forward} \label{sec:forward}

A high significance detection of cosmic polarization rotation would be an important discovery.  While systematic effects currently prevent a definitive measurement of this effect, there are fortunately many avenues to accurately constrain a possible cosmic rotation.

\subsection{Radio Galaxies}
Though popular in the 1990s, the study of the polarization of galaxies in the radio and UV spectrums has been all but abandoned.  There are many improvements to these measurements that can be done.  Simply increasing the number of sources observed should improve the statistical uncertainty on the constraint by $N^{-1/2}$.  Increasing the sample size to $\simeq$400 objects would allow comparable sensitivity to near-future CMB experiments \cite{Kamionkowski2010}.  Measuring the intensity and polarization of radio sources would allow for an $E$- and $B$-mode decomposition.  Computing the intensity-$B$-mode correlation will allow for more sensitive constraints from radio sources (the angular resolution of UV sources is not sufficient to perform $E/B$ decomposition).  A combination of these two improvements would yield significantly higher precision from these extragalactic sources.

CMB measurements have achieved sufficient precision to constrain an overall rotation angle of less than one degree.  Achieving better than one-degree precision from observations of polarized radio and UV galaxies may not be possible.  At these small rotation angles, the ultimate precision will most certainly be limited by the variations in the uniformity of the magnetic fields, and models of their behavior.

\subsection{CMB Polarization Angle Calibration}
In addition to the instruments listed above, there are several CMB polarimeters that will precisely measure the $B$-mode power spectrum. Although these instruments will help constrain an overall polarization rotation, they employ current generation polarization orientation calibration techniques that have limited accuracy.  
However, improved calibrators will provide the accuracy to probe parity violating physics \emph{without} having to build dedicated instruments.  Improvements on current generation calibration technology applied to current and near-future experiments will probe these interesting departures from standard physics with high precision.

The \PB\ and ACTPol constraints are dominated by the $0.5^\circ$ uncertainty of the Tau A measurement \cite{PB2014}.  ALMA \cite{ALMA2009} could measure the polarization of Tau A and Cen A in the same bands to much better precision, allowing a significantly improved constraint on cosmic polarization rotation.  Other polarized astronomical sources, such as 3C274 and 3C58 observed by WMAP \cite{Weiland2011}, can be studied with higher precision instruments.  There are also possible polarized astronomical calibration candidates measured with the Australia Telescope Compact Array (ATCA) \cite{Massardi2013}.  An ideal polarization calibration improvement would be the CalSat CubeSat calibrator described in the following section.

\subsection{CalSat Polarization Calibration Satellite}

CalSat is a proposed dedicated polarization calibration satellite built on the successful CubeSat platform\footnote{www.cubesat.org}.  CalSat will house five linearly polarized microwave sources at 47.1, 80, 140, 249, and 309 GHz (with $< ~1$ MHz bandwidth)\footnote{This creates an all-in-one calibrator for multifrequency experiments, allowing separation of frequency-dependent effects like Faraday rotation}.  CalSat provides bright, well characterized, low cross-polarization polarized microwave radiation in a (polar) low Earth orbit.  CalSat's precession rate will be phased so that it will pass over the entire Earth's surface,  allowing visibility from all ground and balloon-borne observatories.

The CubeSat program was developed by California Polytechnic State University, San Louis Obispo and Stanford University's Space Systems Development Lab as a low-cost space mission option for universities.  CubeSat provides the power, deployment system, integration into the launch vehicle, launch, and telemetry data.  CubeSats are launched as auxiliary payloads on existing missions, significantly reducing the mission cost.

\begin{figure}[h]
\centering
\includegraphics[width=0.49\textwidth]{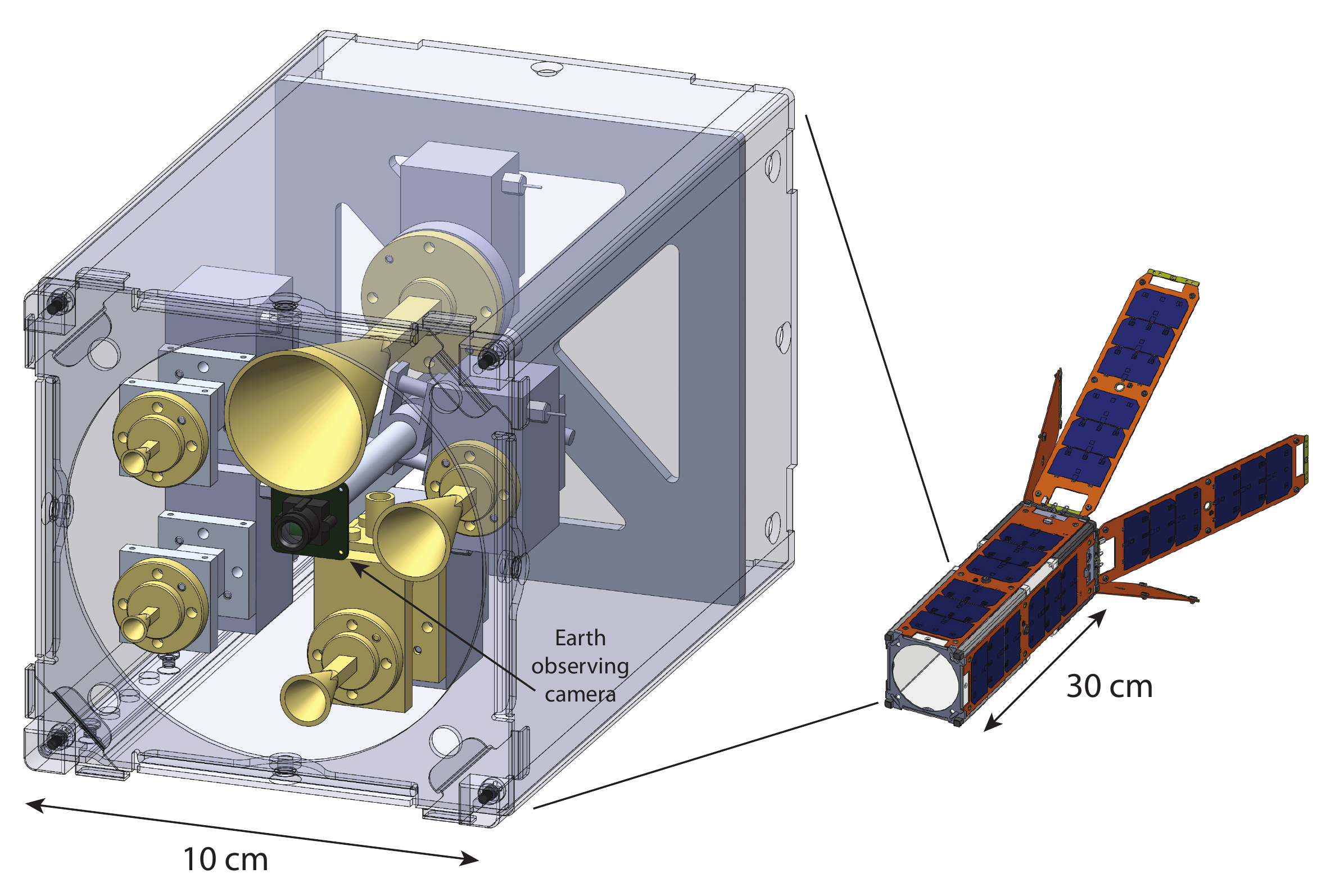}
\includegraphics[width=0.49\textwidth]{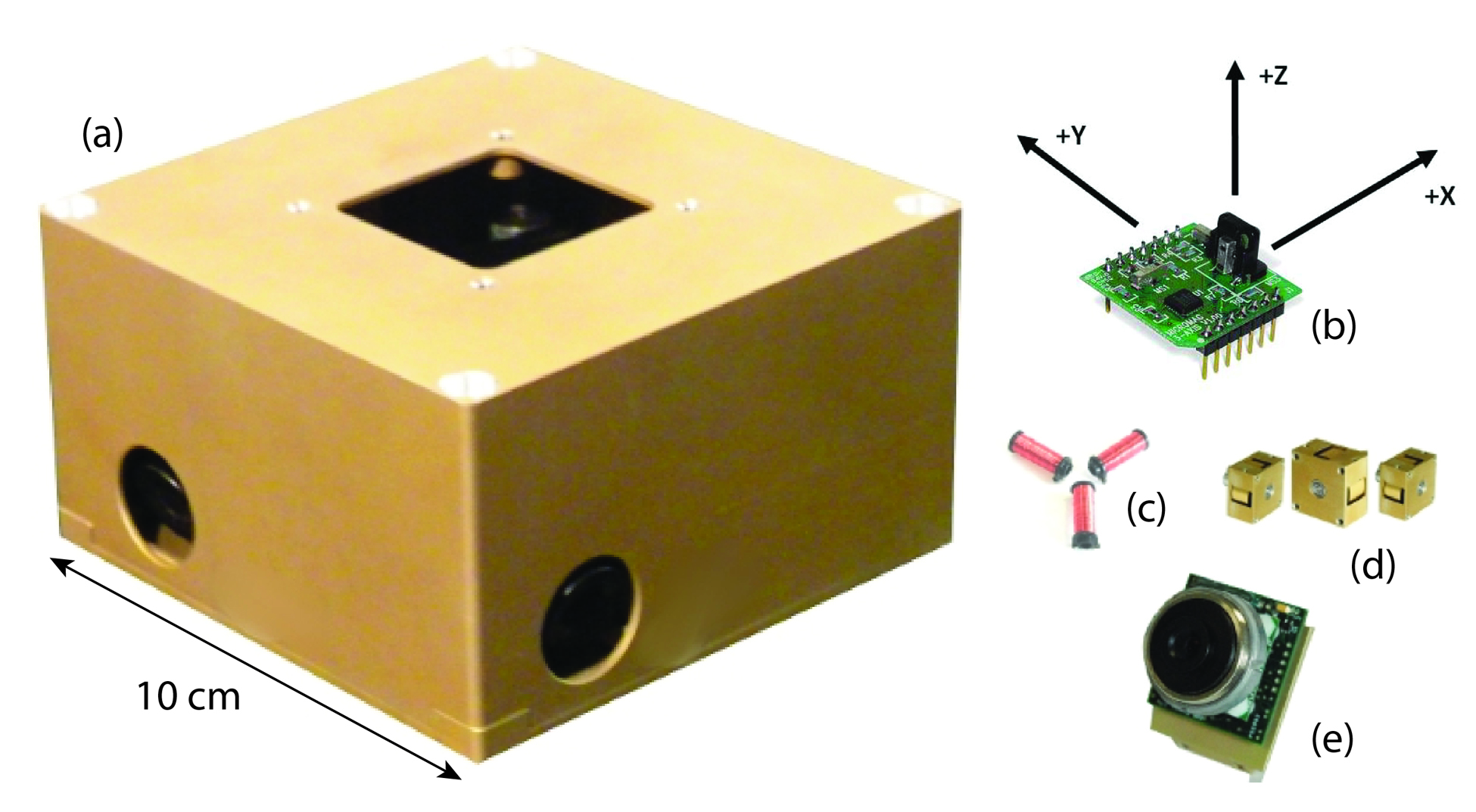}

\caption{\textbf{Left:} A three-dimensional model of the CalSat
  payload. The five horn antennas can be seen inside the circular
  aperture. From largest to smallest, the horns correspond to 47.1,
  80, 140, 249 and 309~GHz. The associated Gunn oscillators and
  multipliers that produce the five frequencies can be
  seen at the back of the horns. A thermal bus connects these
  sources together and to the payload walls. A polarizer will be
  mounted at the payload aperture to ensure polarization purity. A
  control circuit will power and amplitude modulate the
  millimeter-wave sources â each at a slightly different
  frequency. The thermal bus, polarizer and control circuit were
  removed from this figure to show millimeter-wave sources and to show
  clearly that they fit inside the small available volume. \textbf{Right:} The
  attitude determination and control system (ADACS). CalSat will use
  the MAI-400SS ADACS from Maryland Aerospace Inc., which is a turnkey
  system for CubeSats. The CubeSat attitude is measured with a
  magnetometer~(b), star cameras~(e) and Sun sensors, which are
  mounted near the solar panels. The CubeSat attitude is controlled
  with three reaction wheels~(d) and magnetorquer rods~(c), which move
  the CubeSat by torquing against the Earth's magnetic field.}
\label{fig:payload_adacs}
\end{figure}

The CalSat system is composed of three subsystems: the calibrator payload, the spacecraft (shown in Figure \ref{fig:payload_adacs}), and a ground station.  The spacecraft contains the mechanical support structure, power generation systems (including solar panels), the onboard computers and electronics, and the attitude determination and control system (ADACS).  The ADACS includes three reaction wheels, a three-axis magnetometer, two star cameras, three magnetorquer rods, sun sensors and an internal ADACS computer.

The polarization source consists of Gunn diodes with frequencies 47.1 GHz and 80 GHz for the lowest two channels, and a 70 GHz diode with a passive doubler, 83 GHz diode with a passive tripler, and a 103 GHz diode with a passive tripler to generate the 140, 249, and 309 GHz bands respectively.  These bands are designed to be well-matched to the observation windows in the Earth's atmospheric transmittance spectrum, the spectral bands commonly used in polarimeters by the CMB community, and the Amateur Satellite Service bands in the Table of Frequency Allocations used by the Federal Communications Commission (FCC)\footnote{http://www.fcc.gov} and the National Telecommunications and Information Administration (NTIA)\footnote{The United States Frequency Allocation Chart: http://www.ntia.doc.gov/files/ntia/publications/2003-allochrt.pdf}.  The Gunn oscillators couple to rectangular waveguides and produce cross-polarization below the -30 dB level, which will be reduced to -60 dB by the inclusion of a wide-band polarizing grid.

\begin{figure}[t]
\centering

\includegraphics[height=2.5in]{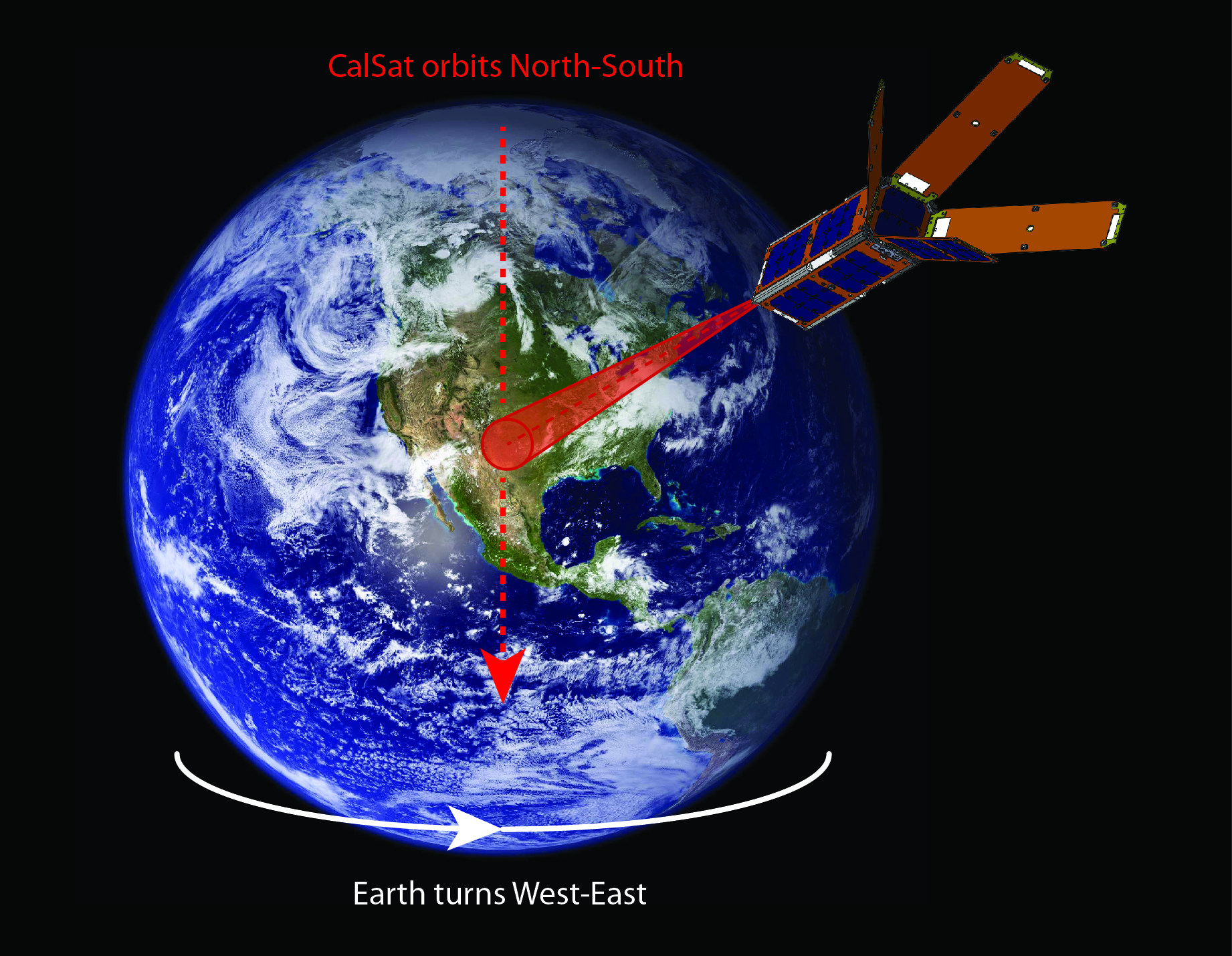}

\caption{A conceptual diagram of the CalSat orbit and
  operation.  For clarity, only one of the five beams is shown.  The
  calibration beam is pointed at the Earth, and the spacecraft is in
  a low-Earth, polar orbit traveling north-south.  The Earth rotates
  below CalSat.  If the precession rate of the orbit is
  phased appropriately, a satellite in a polar orbit like this one
  will, over time, pass over the entire surface of the Earth.
  }
\label{fig:operation}
\end{figure}

Using CalSat's onboard ADACS, the calibrator will be pointed towards CMB receivers with a pointing precision of $< ~1^\circ$.  The polarization axis will rotate as CalSat moves across the sky, providing polarization modulation.  The sources will also be chopped between ``on" and ``off" to mitigate $1/f$ noise from the instrument and atmosphere.   Star camera observations ensure that this rotating polarization axis will always be known to 0.05$^\circ$ precision.  Telemetry data will be made available online for data analysis.  A conceptual diagram of the orbit is shown in Figure \ref{fig:operation}. CalSat will be an order of magnitude improvement over Tau A, the most widely used celestial polarization calibrator.  A comparison of accuracies is shown in Table \ref{table:CalSatvTauA}.

\begin{table}[h]
\caption{Absolute polarization angle systematic uncertainties for Tau A and CalSat.}
\centering
\begin{tabular}{|c|c|c|}
\hline  & 90 GHz & 150 GHz \\
\hline\hline Tau A & $0.5^\circ$ & $0.5^\circ$ \cite{Aumont2009}\\
\hline CalSat & $0.05^\circ$ & $0.05^\circ$ \\
\hline
\end{tabular}
\label{table:CalSatvTauA}
\end{table}

\section{Conclusions} \label{sec:conclusions}

CMB polarimeters are now measuring the minute $B$-mode polarization fluctuations at the tens of nK level.  With these fantastically sensitive instruments we can also constrain cosmic polarization rotation, but only if polarization angle calibration is improved to far-greater accuracy than what is needed to measure the $B$-mode power spectrum.  Radio galaxy and quasar measurements, as well as CMB polarization measurements have been used to claim evidence for cosmic polarization rotation, however galaxy polarization measurements lack the sensitivity to definitively measure the effect and CMB polarimeters are limited by systematic errors on their polarization angle calibration.  No meaningful co-addition of these CMB data sets can be accomplished until at least one experiment can reduce systematic uncertainty on polarization angle calibration to levels subdominant to statistical uncertainty.

In this paper we propose a three-pronged, parallel approach to constrain a possible cosmic polarization rotation signal.  First, better characterization of polarized extragalactic sources is needed.  These measurements complement the CMB measurements by allowing separation of any late-time polarization rotation from early-time polarization rotation, e.g. from effects such as Quintessence \cite{Carroll1998}.  Since these methods may ultimately be limited by intrinsic model and morphology uncertainties, to definitively detect cosmic polarization rotation the CMB represents the most sensitive avenue.

Second, we propose expanding the study of polarized millimeter-wave astrophysical sources by employing instruments such as ALMA. These measurements will help both by improving the precision of current sources, as well as characterizing sources such as those identified by WMAP and ATCA.

Finally, we propose the CalSat mission.  With systematic uncertainty of 0.05$^\circ$ at multiple frequencies,  CalSat would allow current generation CMB experiments to constrain a polarization rotation of order one degree to greater than $15\sigma$ significance. Concomitantly, this polarization angle calibrator can be used not only for measuring cosmic polarization rotation, but also for constraining primordial magnetic fields \cite{Yadav2013}, anisotropic cosmic birefringence \cite{Gluscevic2009}, and for characterization of inflationary $B$-modes.


\section*{Acknowledgments}
The authors wish to acknowledge insightful conversations with Sperello di Serego Alighieri, Mike Niemack, Hans Paar, Meir Shimon, and Edward Wollack.  We also wish to acknowledge our \bicep1, \bicep2, and \PB\ collaborators for many helpful discussions.

\begin{multicols}{2}
\bibliographystyle{h-physrev}
\bibliography{cb_references}
\end{multicols}

\end{document}